\documentclass[%
 reprint,
superscriptaddress,
 amsmath,amssymb,
 aps,
]{revtex4-2}
\usepackage{comment}
\usepackage{graphicx}
\usepackage{xcolor}
\usepackage{dcolumn}
\usepackage{bm}
\usepackage{braket}

\usepackage{todonotes}

\begin{document}

\title{Edge and corner superconductivity in a 2D topological model }

\author{Ying Wang}
\affiliation{Department of Chemistry, University of Southern California, Los Angeles, CA 90089, USA}

\author{Gautam Rai}
\affiliation{Department of Physics and Astronomy, University of Southern California, Los Angeles, CA 90089, USA}
\affiliation{I. Institut f\"{u}r Theoretische Physik, Universit\"{a}t Hamburg, 22607 Hamburg, Germany}

\author{Stephan Haas}
\affiliation{Department of Physics and Astronomy, University of Southern California, Los Angeles, CA 90089, USA}

\author{Anuradha Jagannathan}
\affiliation{
Laboratoire de Physique des Solides, B\^at.510 Universit\'{e} Paris-Saclay, 91400 Orsay, France
}

\begin{abstract}
We consider a two-dimensional generalization of the Su-Schrieffer-Heeger model which is known to possess a non-trivial topological band structure. For this model, which is characterized by a single parameter, the hopping ratio $0 \leq r\leq 1$, the inhomogeneous superconducting phases induced by an attractive $U$ Hubbard interaction are studied using mean field theory. We show, analytically and by numerical diagonalization, that in lattices with open boundaries, phases with enhanced superconducting order on the corners or the edges can appear, depending on the filling. For finite samples at half filling, the corner site superconducting transition temperature can be much larger than that of the bulk. A novel proximity effect thus arises for $T_{c,bulk} < T<T_{c,corner}$, in which the corner site creates a nonzero tail of the superconducting order in the bulk. We show that such tails should be observable for a range of $r$ and $U$ values.
\end{abstract}

\maketitle

\section{Introduction}
In this paper, we consider the effects of attractive on-site interactions in a two-dimensional Su-Schrieffer-Heeger (2D SSH) model. This model is an extension of the well-known SSH chain \cite{su1979solitons,su1983erratum} to two dimensions, with alternating weak ($t_1$) and strong ($t_2$) bonds along both spatial directions. In finite lattices, the non-interacting model has edge and corner modes in the appropriate topological sectors and is an example of higher order topological insulators (HOTI)  \cite{trifunovic2021higher}. Using the Bogoliubov-de Gennes formalism, we show that adding a Hubbard-type attractive on-site interaction results in a variety of inhomogeneous superconducting states in which the pairing is site-dependent. Fig.~\ref{deltapattern} shows examples of the possible different kinds of low temperature superconducting phases---one can obtain phases for which the pairing order (a) is quasi one-dimensional, essentially restricted to the edges, (b) is largest in the bulk and very small on the periphery, or (c) is enhanced on the corner sites. The spatial and thermal properties of these phases depend on the Hubbard interaction strength as well as on the ratio of hopping amplitudes of the model, $r=t_1/t_2$.

HOTI have been much studied recently. The formation of electric multipole moments and charge pumping in such a lattice has been addressed \cite{benalcazar2017electric,benalcazar2017quantized}. Photonic systems based on the 2D SSH model were investigated in \cite{xu2020general}. A classification scheme for topological superconductors and the bulk-boundary correspondence in Bogoliubov-de Gennes type models has been discussed in \cite{trifunovic2021higher,geier2020symmetry,luo2022higher}. In this paper we consider only the simplest possibility of $s$-wave pairing, however, we expect that similarly interesting edge and corner phenomena should appear when the basic model is extended to permit other types of superconducting pairing. The present study constitutes, for example, a good starting point for investigations of corner and edge Majorana fermions. 

While edge modes and resulting higher order topological superconducting phases have been reported before in the literature, as on the honeycomb lattice \cite{scammell2022intrinsic,li2020artificial},  the present model is of particular interest since it provides an analytically tractable example with a tunable parameter, whose ground state and finite temperature properties can be described in detail. Furthermore, we observe an interesting interplay between surface and bulk superconductivity which has not been reported in HOTI structures so far.

This paper is organized as follows. In Sec.~II we introduce the model, and the Bogoliubov-de Gennes (BdG) mean field approach used here.  In Sec.~III we present results obtained by numerical diagonalization in 2D SSH systems. The dependence of local pairing order parameters on interaction strength, hopping ratio and chemical potential are described. The critical temperatures of corner and bulk superconductivity are obtained. For a range of temperatures, a mixed state is shown to exist in OBC systems. Sec.~IV presents theoretical analysis, starting with the solution for systems with periodic boundary conditions and then its generalization to topologically non-trivial finite systems. We conclude in Sec.~V with a discussion and perspectives for future work.

\section{Attractive Hubbard model on the 2D-SSH lattice}
\subsection{The noninteracting Hamiltonian }

\begin{figure}
    \includegraphics[width=\columnwidth, trim = 0  0 160pt 180pt, clip]{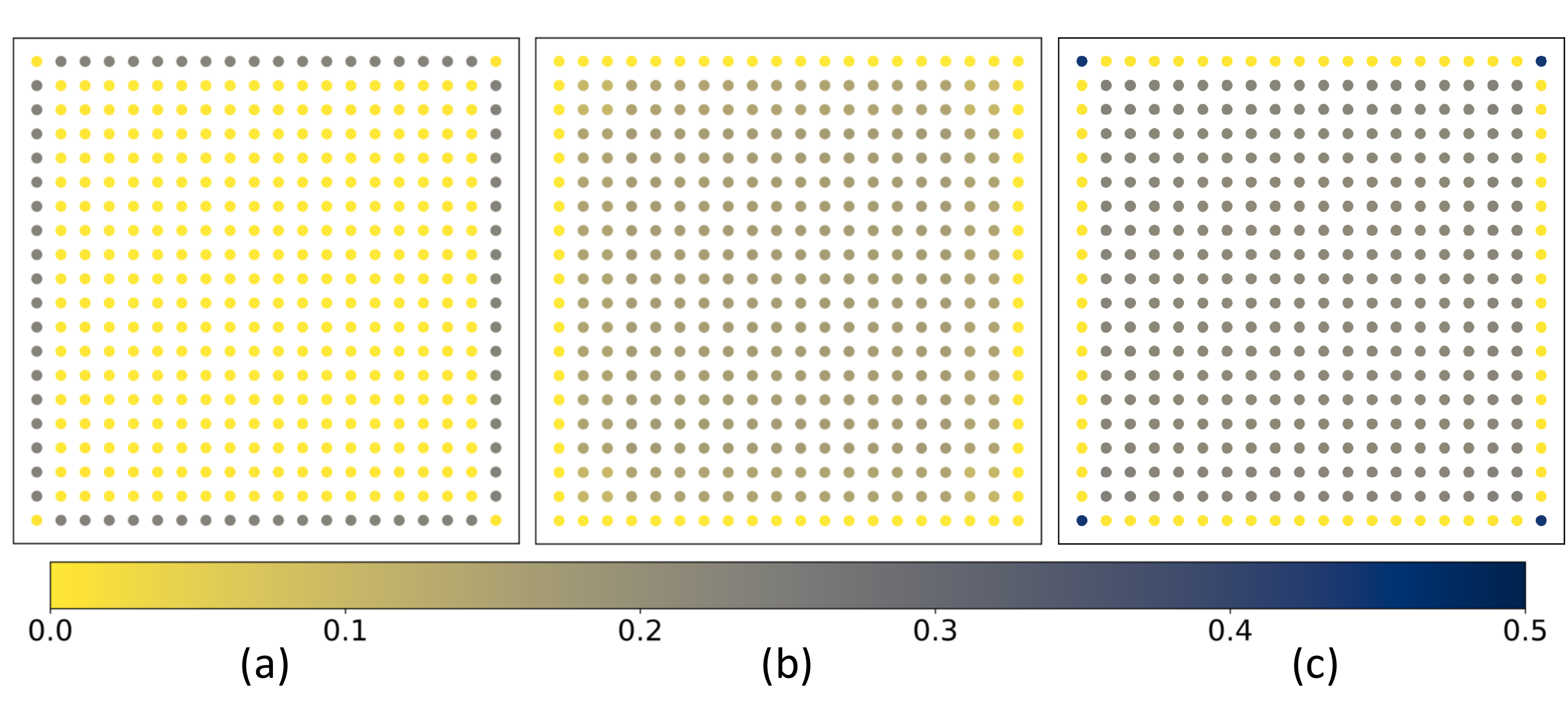}
    \caption{Spatial variation of the local superconducting order parameters in a finite 2D Su-Schrieffer-Heeger lattice for different band fillings (a) $1/4$-filling (in the edge band), (b) $0.362$-filling (in the central bulk band) and (c) $1/2$-filling. Parameters used: $r=0.1$ and $V=1$.}
    \label{deltapattern}
\end{figure}

\begin{figure}
    \includegraphics[width=1\columnwidth, trim = 0  0 320pt 90pt, clip]{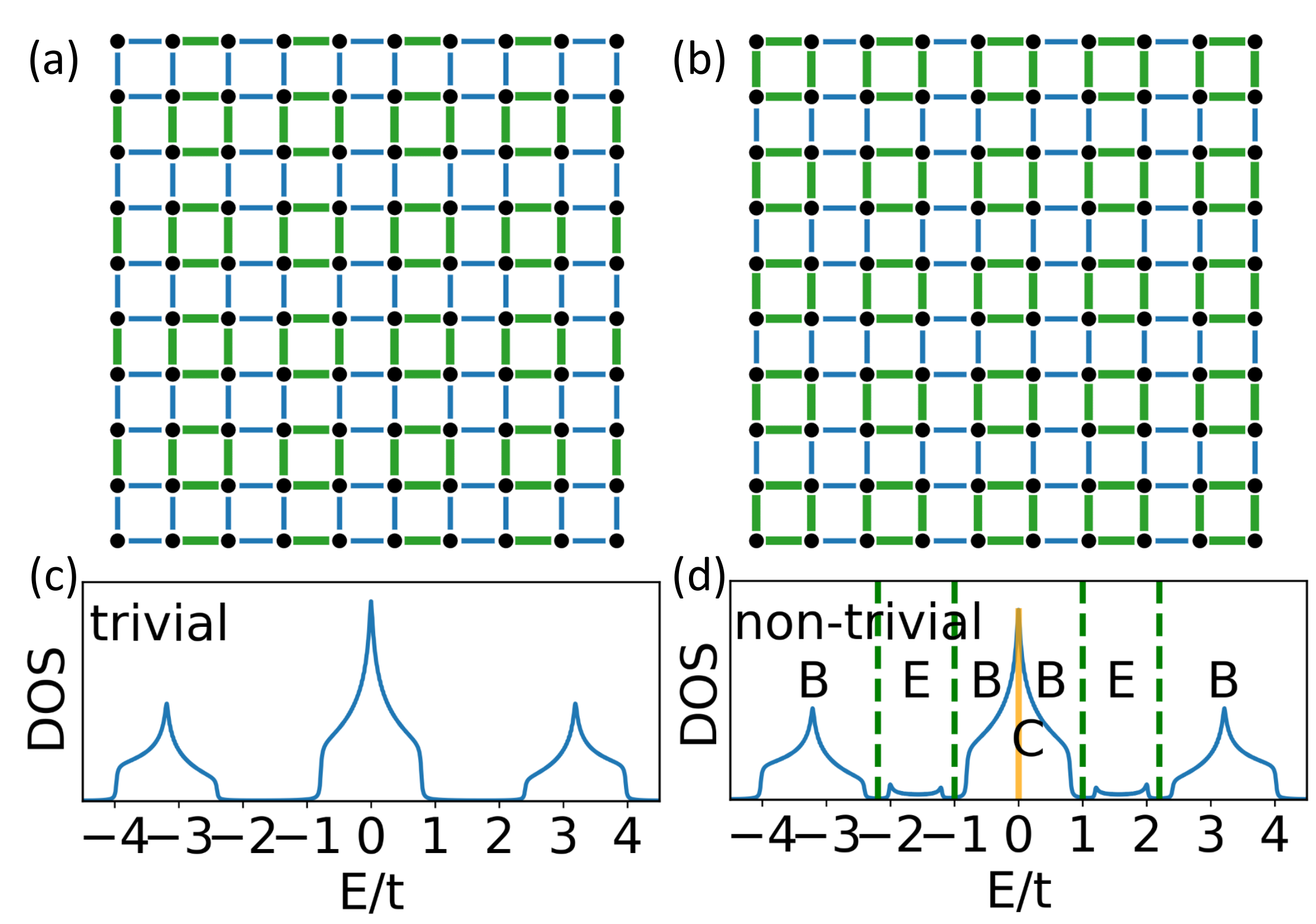}
    \caption{Two finite systems showing a) a non-trivial case b) and the trivial case (black dots represent lattice site, $t_1$ is represented by blue thinner bond, $t_2$ is green thicker bond).  (c) DOS for the OBC trivial lattice. PBC lattice has the same DOS as the OBC trivial lattice when the latter has enough unit cells. (d) DOS for the OBC non-trivial case. Bands in this DOS plot are labelled by B (bulk), E (edge) and the peak at $E=0$ includes a contribution C (corner). In these calculations, $r=0.25$.}
    \label{latticeandDOS}
\end{figure}

The 2D-SSH tight-binding model is defined for sites lying on vertices of a square lattice of side $a$, and involves two different nearest neighbor hopping amplitudes, $t_1$ and $t_2$. Along the horizontal ($x$) direction, the sequence of hopping amplitudes is alternating, just as in the parent 1D SSH model \cite{su1979solitons, su1983erratum}. The same is true for the hopping along vertical ($y$) direction, as shown in Fig.~\ref{latticeandDOS} (a) and (b).  We will assume $t_1 \leq t_2$ and discuss properties of this model as a function of the hopping ratio $r=t_1/t_2 \leq 1$. The hopping amplitudes are furthermore assumed to be positive (since the sign changes can be gauged away).  The non-interacting Hamiltonian is then
\begin{equation}
    \label{sshham}
H_{0} = -\sum_{\langle i,j\rangle} t_{ij} c^\dagger_{i\sigma}c_{j\sigma} 
        + h.c.
\end{equation}

where $\langle i,j\rangle $ denote nearest neighbor sites $i$ and $j$, while $\sigma$ denotes spin. The unit cell consists of four sites. For the infinite system or finite systems with periodic boundary conditions (PBC), one can diagonalize by Fourier transforming.  Note that the energy spectrum can be obtained very easily since the problem is separable in $x$ and $y$ variables, giving rise to two 1D SSH spectra. The 2D spectrum is just the direct sum of the energy bands of the 1D SSH model, $\epsilon_{1D}(k_x)$ and $\epsilon_{1D}(k_y)$. Wavefunctions are products of the 1D SSH wavefunctions, that is, $\psi(k_x,k_y)=\psi_{1D}(k_x)\psi_{1D}(k_y)$. The corresponding energy dispersion relations are given by

\begin{eqnarray}
    \label{eq:spectrum}
    \epsilon^{(n)}_{2D}(k_x,k_y) =  \pm \epsilon_{1D}(k_x) \pm \epsilon_{1D}(k_y) \\ 
   \epsilon_{1D}(k) = t_2\sqrt{1+r^2 + 2r \cos 2k a} 
\end{eqnarray}

where the wave vectors lie in the Brillouin zone, $-\pi/2a \leq k_j \leq \pi/2a$ ($j=x,y$). The parameter determining the spectral properties is $r$, while $t_2$ serves only to set the global scale of energy. The four choices of sign in the above expression for $\epsilon_{2D}$ $(++, +-,-+,--)$ correspond to the 4 bands labeled $n=1,..,4$. The two overlapping central bands ($n=2$ and $3$) intersect along the diagonals, where $\epsilon^{(2)}_{2D} =\epsilon^{(3)}_{2D}=0$. They transform into each other under the mirror symmetries that exchange  $k_x \leftrightarrow \pm k_y$.
The energy spectrum is particle-hole symmetric as the model is bipartite. When $r<\frac{1}{2}$, a gap separates the lateral bands from the central bands. There are logarithmic van Hove singularities at $\epsilon=0$, and at the centers of the two lateral bands. 

With open boundary conditions (OBC), topologically protected states can arise at the edges. We use the term ``weak edge'' when all sites lying on the edge are connected to the interior by weak bonds. This configuration results in the appearance of 1D edge modes, which have wave modulations along the edge, but decay exponentially along the direction perpendicular to the edge.  Where two weak edges meet, there is an additional ``zero-dimensional'' corner mode, which is exponentially decaying in both directions. The localization length $\xi$ is that of the edge states of the 1D SSH chain and depends on the hopping ratio, $\xi=2a/ |\ln r|$. A precursor of this 2D lattice, a ladder type system of 2 coupled chains has been studied in \cite{padavic2018topological}. 

The total density of states (DOS) of the finite trivial and non-trivial 2D SSH lattice is shown in  Fig.~\ref{latticeandDOS} (c) and (d). The nontrivial system has two supplementary bands corresponding to the quasi 1D edge modes. The spectrum has four gaps for small $r$, which close when $r=1/3$.  If present, each of the zero-dimensional modes localized on corners of the square contributes a delta function $\delta(E)$ to the density of states.  

In the interacting problem, an attractive onsite Hubbard term, $H_{int}$,  is added to $H_{0}$ with
\begin{align}
    \label{sshint}
    H_{int}= -V \sum_i \hat{n}_{i\sigma} \hat{n}_{i\bar{\sigma}}
\end{align}
($V>0$), where $\hat{n}_{i\sigma}=c^\dagger_{i\sigma}c_{i\sigma}$ is the number of electrons of spin $\sigma$ on the site $i$ and $\bar{\sigma}$ represents the opposite spin of $\sigma$. We assume that the instability of interest is the $s$-wave superconducting instability. In particular, at half-filling, a competing charge density wave instability could exist but be suppressed by doping or adding a small next nearest neighbor hopping.   
To proceed, we use a standard mean field approximation to write the following effective total Hamiltonian 
\begin{align}
\label{eq:MF-Hamiltonian}
    H_{BdG} &= \sum_{i\sigma} (u_i^{HF}-\mu_i)c^\dagger_{i\sigma} c_{i\sigma} - \nonumber \\ 
    \sum_\sigma \sum_{\langle i,j\rangle} &t_{ij} (c^\dagger_{i\sigma}c_{j \sigma} 
        + h.c.)+ \sum_{i}  (\Delta_{i}c_{i\downarrow}c_{i\uparrow}  + h.c.) 
\end{align}
where $\mu_i$ is the chemical potential. The mean fields are the Hartree-Fock shift $u_i^{HF} = V\langle c^\dag_{i\uparrow}c_{i\uparrow}\rangle$ and the (real) local superconducting order parameters (OP) $\Delta_i =V\langle c_{i\downarrow}c_{i\uparrow}\rangle$. These quantities are determined self-consistently for different choices of band filling and boundary conditions. They are site-independent in the infinite lattice, and in finite systems with periodic boundary conditions, but become site-dependent when there are edges.   

\begin{figure}[htb]
    \includegraphics[width=\columnwidth, trim = 0  0 385pt 270pt, clip]{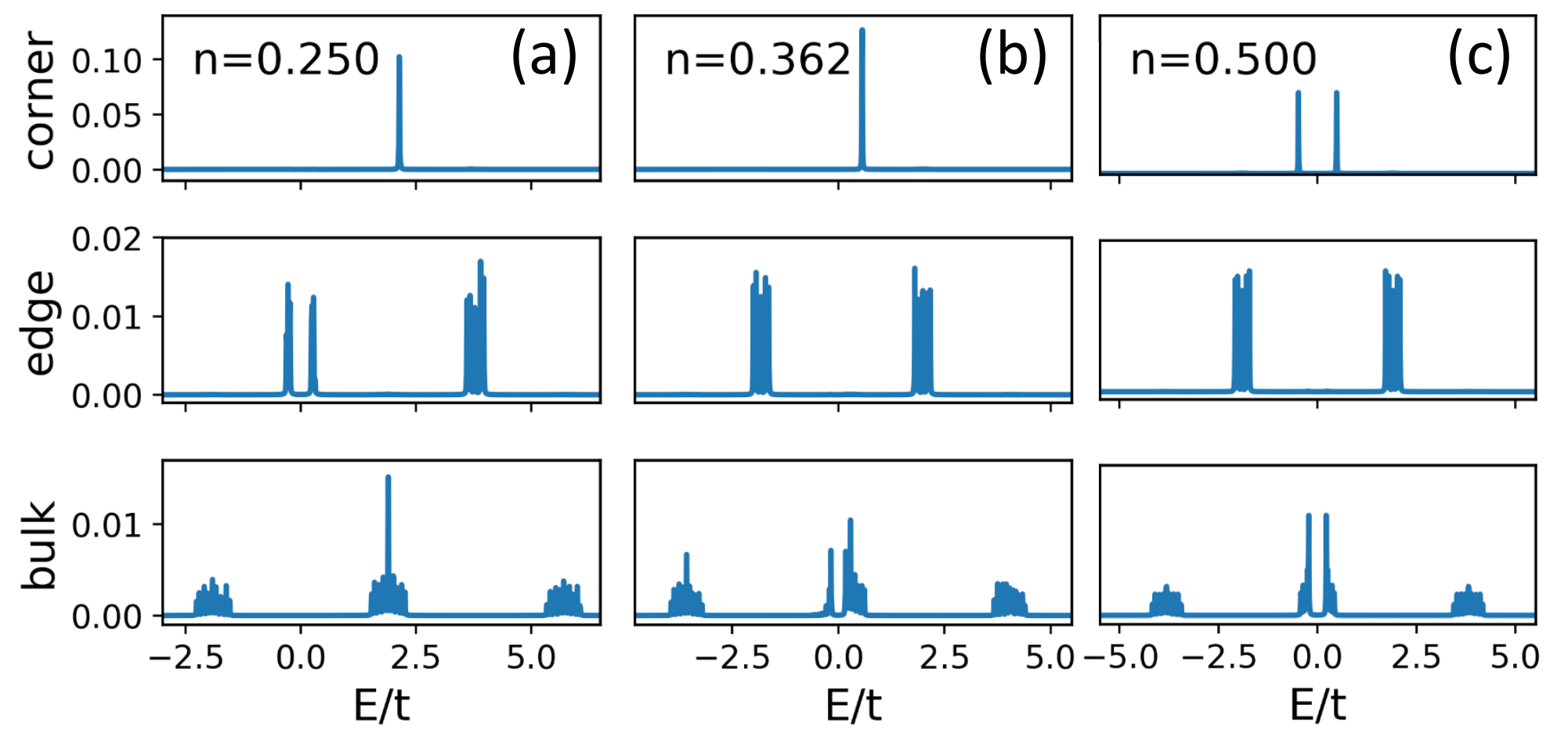}
    \caption{Plots of the LDOS of corner, edge, and bulk sites  for different band fillings. Parameters used: $r=0.1$ and $V=1$.}
    \label{LDOS}
\end{figure}

\section{Numerical Results}
\begin{figure}[htb]
    \includegraphics[width=\columnwidth, trim = 10pt  0 170pt 200pt, clip]{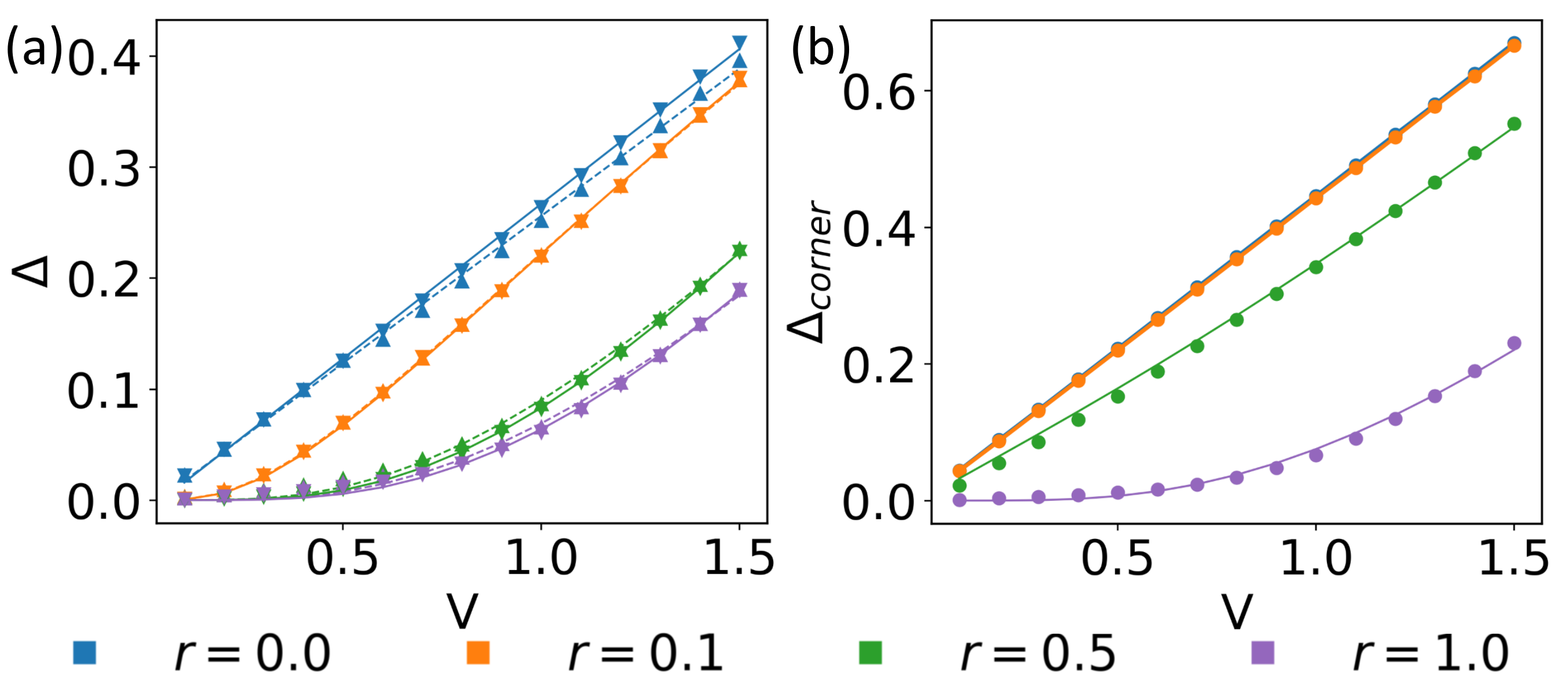}
    \caption{Plots of the $T=0$ order parameters versus $V$ with fitting curves. (a) $\Delta$ for PBC (up-triangle with dash line), $\Delta_{bulk}$ for OBC (down-triangle with solid line); (b) $\Delta_{corner}$ for OBC (circle with solid line). Points indicate numerical results, and the continuous lines are fits to the analytical expressions (see text).}
    \label{deltaVfitting}
\end{figure}

\begin{figure}[htbp]
    \includegraphics[width=\columnwidth, trim = 0 0 175pt 220pt, clip]{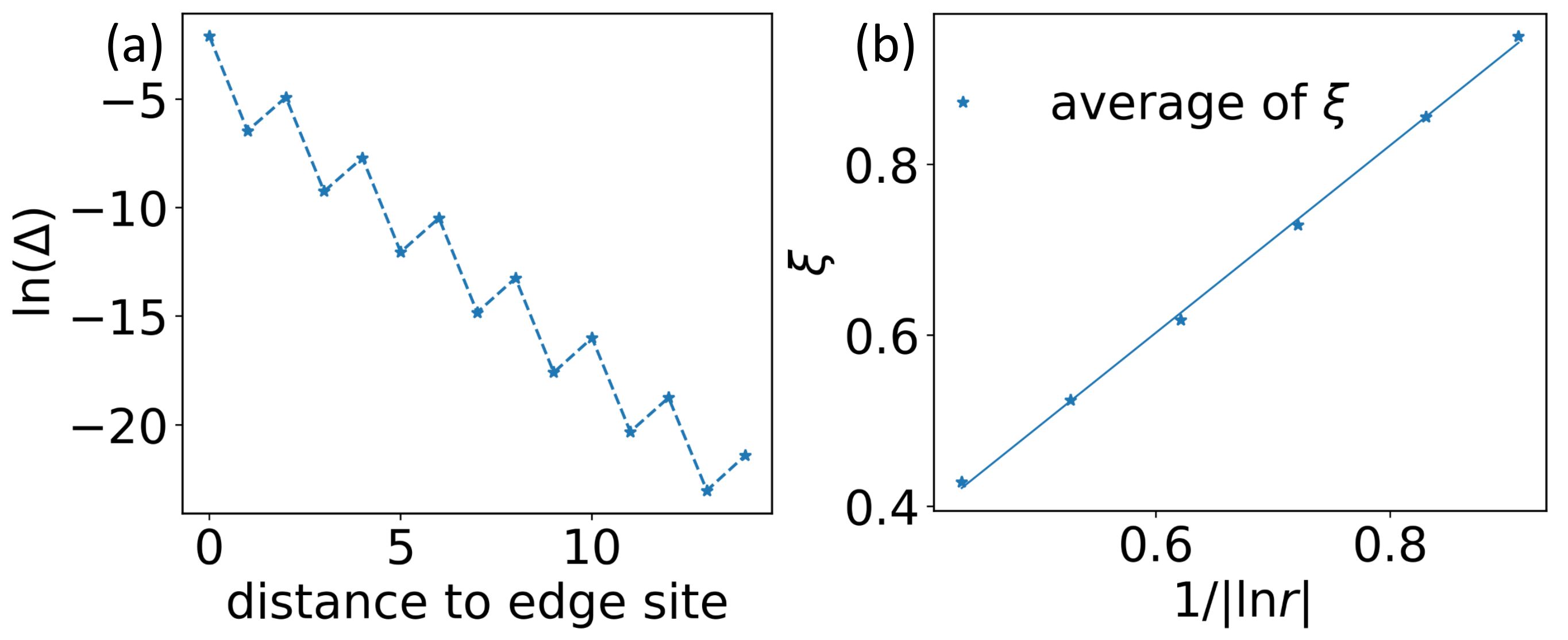}
\caption{(a) Plot of $\ln \Delta_{edge}$ versus distance to the edge under $V=1$ and $r=0.25$, showing the exponential decay and the odd-even oscillation (see text). (b) Plot of penetration length $\xi$ as a function of $1/ | \ln{r}|$. The points show the values of $\xi$ under differet ratios of r, while the line shows the expected theoretical dependence. $r=0.1,0.15,0.2,0.25,0.3,0.333$ from left to right.}
\label{edgepenetration}
\end{figure}

\subsection{Order parameter results}

We now present numerical solutions of the BdG equations for finite 2D-SSH samples. The real space method used to solve Eq.~\ref{eq:MF-Hamiltonian}, was originally developed to study the inhomogeneous superconducting state in disordered systems \cite{ghosal1998role}. It has also been applied to study periodic systems such as the checkerboard Hubbard model \cite{aryanpour2006effect}.
The calculation determines the pairing parameter $\Delta_i$ for each site self-consistently as follows: an initial ansatz is made for the BdG Hamiltonian using randomly chosen values for the OPs, $\Delta^{(0)}_i$. The Hamiltonian is diagonalized numerically, to find the eigenvalues $E_n$ and corresponding eigenvectors $\{u_n, v_n\}$. New values of $\Delta_i$ are computed using the expression   
\begin{align}
    \Delta^{(1)}_{i} = V_{i}\sum_{n} v_{in}^{*}u_{in}\Big[1-2f(E_{n}, T)\Big]
\end{align}
where $f(E_n, T)$ is the Fermi-Dirac distribution.
These $\Delta^{(1)}_i$ are injected back into the BdG hamiltonian, and the calculation is iterated until convergence is reached. The calculations have been done under a fixed bandwidth condition, that is, we vary the hopping ratio $r$ keeping the bandwidth $W=2(t_1+t_2)=4t$ constant. Results are reported in units of $t$, the average hopping amplitude.

In Fig.~\ref{deltapattern} we have illustrated the local OP distribution for different band fillings, in a $20\times 20$ lattice subjected to OBC, with $r=0.1$ and $V=1$. Nb. in all the open systems considered, all four edges are taken to be weak edges. Panel (a) corresponds to the edge band being half-filled ($n=0.25, \mu = 1.90t$). Here the OP is largest on the edges and decays exponentially into the interior as we will show later. In panel (b), the central bulk band is partially filled ($n=0.362, \mu = 0.19t$). The OP is accordingly largest in the bulk of the sample. In panel (c), the system is at half-filling ($n=0.5, \mu = 0$). Here the OP is strongest on the four corner sites, followed by the bulk, while the edges have negligibly small OP. The local densities of states (LDOS) for these three band fillings on the corner, edge, and bulk sites are shown in Fig.~\ref{LDOS}. These plots rationalize the site-dependent superconductivity pattern. The gap in LDOS is seen only at the edge sites in (a), the bulk sites in (b), and both corner and bulk sites in (c). The most interesting situations, corresponding to cases a) and c), are discussed below.

\bigskip

\noindent {\it Chemical potential at the band center (half-filling)}

In Fig.~\ref{deltaVfitting}, we plot $T=0$ order parameters as a function of $V$ for several values of the hopping ratio $r$. The system sizes are large enough that the results have converged, and the error bars are less than the size of the symbols used in the plots. Panel (a) shows the order parameter in the center of the sample for OBC, labeled $\Delta_{bulk}$, compared to the order parameter for PBC, labeled $\Delta$. At half-filling, we note that, as the hopping ratio $r$ increases from 0 to 1, the bulk OP decreases and becomes smallest for $r=1$. Panel (b) shows the order parameter at the corner sites. Figs.~\ref{deltaVfitting} show that, when $r=0$, all of the order parameters are proportional to $V$. However for non-zero $r$, the bulk order parameters vary as $\exp^{-\sqrt{cst/V}}$, as obtained for the half-filled square lattice \cite{hirsch1985two}. This type of scaling with $V$ is expected when the Fermi level is located at a logarithmic van Hove singularity \cite{noda2015bcs,noda2015magnetism}. The corner OP is well-fitted by an extrapolation between the linear and exponential terms as follows 
\begin{align}
    f(V) = c_1 V + c_2 \exp^{-\sqrt{c_3/V}}
\label{linearandexpfittingformula}
\end{align}

Values of the fitted constants are given in Table.~\ref{table-delta-V-fitting} for each of the OP. Note that $c_2$ vanishes when $r\rightarrow 0$, so that the variation is purely linear in V in this limit, while for $r=1$, the linear term vanishes. These behaviors will be explained in Sec.~IV.

\begin{figure}
    \includegraphics[width=0.8\columnwidth, trim = 0 0 695pt 300pt, clip]{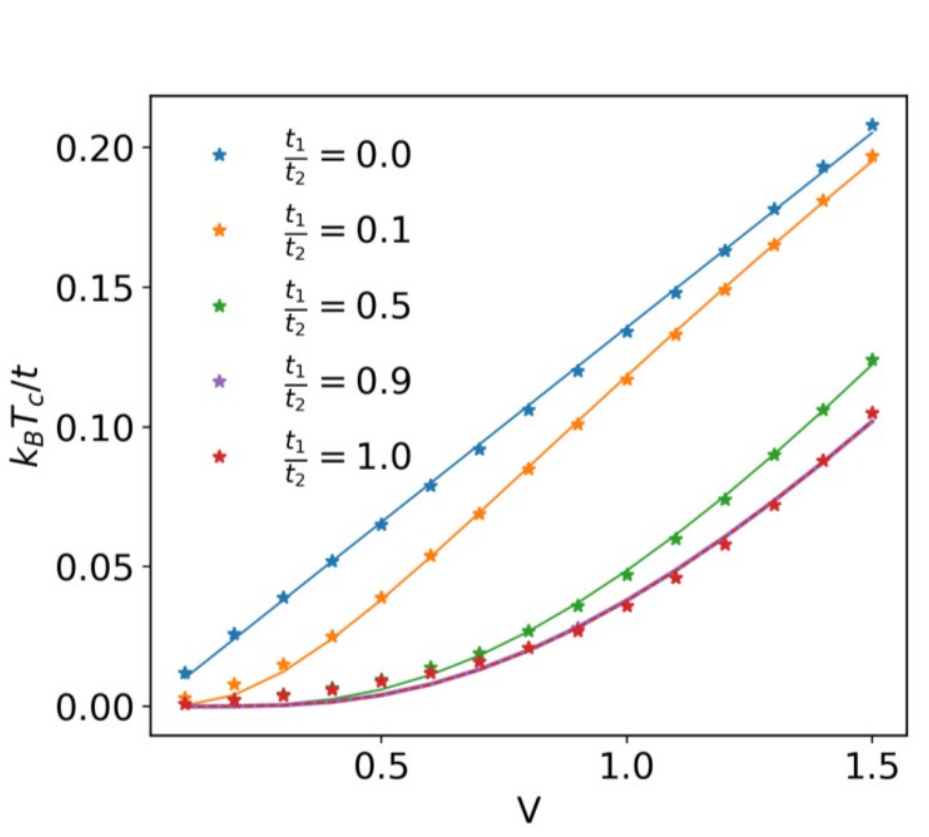}
    \caption{PBC critical temperatures plotted vs. V for ratio values of 0.0, 0.1, 0.5, 0.9, 1.0.}
    \label{PBCTcVfitting}
\end{figure}  

\begin{figure}
    \includegraphics[width=\columnwidth, trim = 0 0 400pt 200pt, clip]{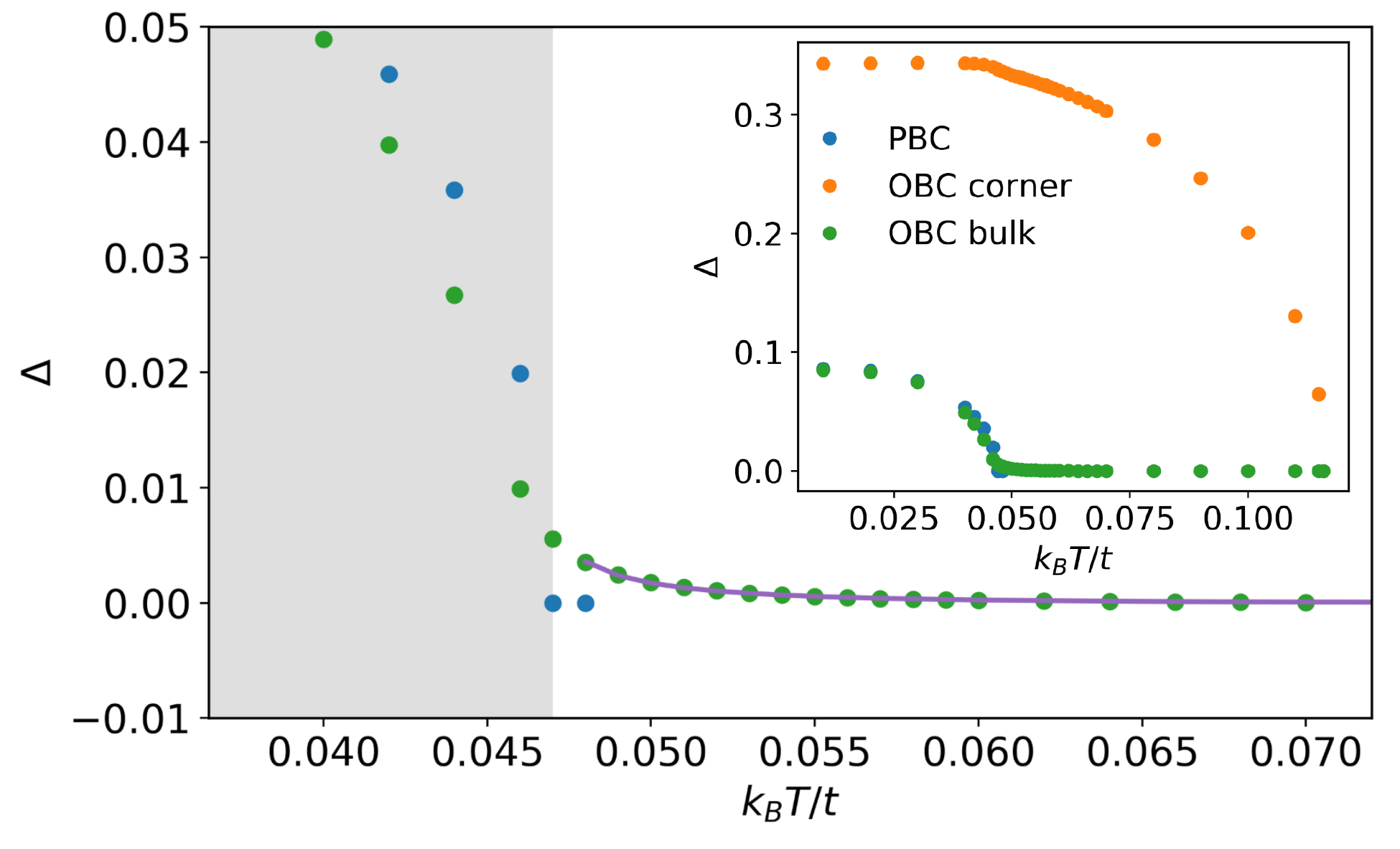}
    \caption{Inset: $T$-dependence of $\Delta_{corner}$ and $\Delta_{bulk}$ for OBC and of $\Delta$ for $r=0.5$ and $V=1$ lattice. The main figure shows $\Delta_{bulk}$ versus $T$. The low  temperature region $T<T_{c,PBC}$ is colored grey, and the tail of the bulk OP has been fitted to the expression given in the text.}
    \label{bulkTcfitting}
\end{figure}

\bigskip

\noindent {\it Chemical potential in the edge band }

When the chemical potential lies within the edge band (i.e. just above $\frac{1}{4}$-filling), a superconducting gap is opened in the edge band, as can be seen in Fig.~\ref{LDOS} (a).  The spatial dependence of the order parameter is governed by the spatial properties of the 1D SSH edge modes, which are well-known. One thus observes, in addition to the two-sublattice structure, an exponential decay as a function of the distance from the edge. These features are shown in Fig.~\ref{edgepenetration}(a) which presents a log-linear plot of $\Delta_{edge}$ versus the distance. Fig.~\ref{edgepenetration}(b) shows the values of the fitted localization length $\xi$ (points) as a function of the hopping ratio $r$, along with the expected dependence given by $\xi=a/\vert\ln(r)\vert$ (line). Finally, $\Delta_{edge}$ has the standard BCS dependence on $V$, namely $\Delta_{edge} \sim \exp^{-cst/V}$.

\subsection{Finite temperature results}
Fig.~\ref{PBCTcVfitting} shows results for the critical temperature of the periodic model at half-filling, $T_{c,PBC}$, plotted against $V$, for different choices of the hopping ratio. Lines are fits to the data using the form $\exp^{-\sqrt{cst/V}}$.  

For OBC, and a range of $r$-values, we find distinct transition temperatures for the bulk and corner OP. This kind of two-step transition with corner (or surface) superconductivity, followed by bulk superconductivity can be found more generally, in systems with boundaries surface states, as shown in \cite{samoilenka2020boundary,samoilenka2020pair,benfenati2021boundary}.
The temperature-dependence of the order parameter at bulk and corner sites, as well as for PBC, are shown in Fig.~\ref{bulkTcfitting} for $r=0.5$ and $V=1$. Fig.~\ref{bulkTcfitting} shows that the corner and bulk sites have different transition temperatures. The bulk OP which is expected to go to 0 at the bulk transition temperature $T_{c,PBC}$, actually shows a non-zero tail above this critical temperature. This tail arises from a proximity effect due to the corner site where the OP is still non-zero, and it accordingly vanishes when $T=T_{c,corner}$. One expects that the tail should be proportional to the corner OP, and also depend on $R$, the distance of the midpoint from the corner as $e^{-R/\lambda(T)}$. Here $\lambda(T)$ is the correlation length. Close to the bulk transition the correlation length should vary as $\lambda(T)=A/\sqrt{T-T_{c,PBC}}$ in mean field theory. In this formula the prefactor $A$ depends on the hopping ratio, increasing monotonically as $r\rightarrow 1$. Indeed, as shown in Fig.~\ref{bulkTcfitting}, the tail is well fitted by this form. The tail is clearly visible only within a range of $r$ values. For very small hopping ratios, when the correlation length $\lambda(T)$ is too short for the tail to be observable. For larger $r \sim 1$, the correlation length is large and corner and bulk critical temperatures very close so that no tail is observed.

\begin{table}[h!]
\begin{center}
\begin{tabular}{|c c c c c c c|} 
\hline
r& &0.0 &0.1 &0.5 & 0.9 &1.0 \\
\hline 
$\Delta_{corner}(0)$ &$c_1$ &0.44 &0.44 &0.32 &0.04 &0 \\
 &$c_2$ &0 &0.02  &4.96 &23.32 &26.27 \\
 &$c_3$ &0 &8.33 &28.96 &34.40 &34.27 \\
\hline 
\end{tabular}
\caption{Values of fitting parameters for zero-temperature corner site order parameters $\Delta_{corner}(0)$ to the function $f(V)$ for different $r$.}
\label{table-delta-V-fitting}
\end{center}
\end{table}

\section{Theoretical analysis}
\subsection{Calculation for periodic boundary conditions}

We begin by considering a translationally invariant 2D SSH square of side $L$ with $N=(L/2a)^2$ unit cells, subject to periodic boundary conditions. Due to the lattice symmetries, all sites have the same order parameter $\Delta_i=\Delta$. We outline the gap equation that is obeyed in the particle-hole symmetric half-filled lattice, $\mu=0$.  We will consider the weak coupling limit, $V$ small, and define the  expectation values $b_{\vec{k}}$ by 
\begin{align}
    \label{eq:bdef}
 b^{(n)}_{\vec{k}} = \langle \eta^{(n)}_{\vec{k}\downarrow}\eta^{(n)}_{-\vec{k}\uparrow}\rangle     
\end{align}
with all other expectation values of two annihilation operators assumed to vanish, by the symmetries of the problem. $\eta^{(n)}_{\vec{k}\sigma}$ are eigenmodes in Fourier representation. The order parameter $\Delta$ can be written in terms of a sum over bands using the transformation to the diagonal basis. One has $\Delta=\sum_n \Delta^{(n)}$ where 
\begin{align}
    \label{eq:gOP}
\Delta^{(n)}=\frac{V}{4N}\sum_{\vec{k}}b^{(n)}_{\vec{k}}
\end{align}
We can additionally simplify by assuming that only the two central bands contribute, and fix the band index at $n=2$ (it suffices to keep only one of the two central bands in the sums over $\vec{k}$, by virtue of their symmetry under exchange of $k_x$ and $k_y$). The BdG equations for different $\vec{k}$ decouple, giving rise to 2 by 2 matrices of the form
\begin{eqnarray}
H_{\vec{k}} = \left (\begin{array}{c c}
\epsilon_{2D}(\vec{k}) & \Delta \\
\Delta & -\epsilon_{2D}(\vec{k}) 
\end{array}\right)
\end{eqnarray}
where $\epsilon_{2D}(\vec{k})$ is given by Eq.~\ref{eq:spectrum}. Diagonalization yields quasiparticle energies of the form $E(\vec{k})=\sqrt{\epsilon^2_{2D}(\vec{k}) + \Delta^2}$.  As in the standard case, the gap equation is obtained from the self-consistency condition, which reads
\begin{align}
    \label{eq:gapeqPBC}
    \Delta(T) 
    = \frac{V}{2}\int d\epsilon ~\rho(\epsilon) \frac{\Delta(T)}{E}\mathrm{th} (\beta E/2)
\end{align}

where $\beta=1/k_BT$ is the inverse temperature and $\epsilon=\epsilon_{2D}$ is the single particle energy. This gap equation predicts that, for fixed $V$, the order parameter $\Delta(0)$ decreases as a function of $r$. In particular, for $r$ small perturbative expansion predicts a decrease of the OP proportional to $r^2$. This is indeed seen in Fig.\ref{deltaVfitting}.

In the limit $r\rightarrow 0$, the gap equation can be solved to obtain the $T=0$ order parameter $\Delta(0)$ as a function of $V$. As the width of the central band tends to zero, the DOS can be approximately replaced by a delta function $A\delta(E)$ where $A\approx 0.5$ is the fraction of states lying within this band (neglecting correction of order 1/L). The gap equation yields $\Delta(0) \sim V$. The critical temperature can be determined from the gap equation from the requirement that $\Delta(T_c)=0$. In the limit of small $r$, $T_c$ scales similarly as the OP, that is, $T_{c} \propto V$. 
  
For non-zero $r$ the integral in Eq.\ref{eq:gapeqPBC} is determined by the logarithmic van Hove singularity at $E=0$. Instead of the standard BCS form, $T_c \propto \exp^{-1/N_0V}$, that is expected for a regular density of states (where $N_0$ is the DOS at the Fermi level), the critical temperature here has a $V$-dependence given by $T_c \sim  \exp^{-\sqrt{cst/V}}$ \cite{noda2015bcs,noda2015magnetism}. These behaviors are confirmed by the numerical calculations, as shown in Fig. \ref{PBCTcVfitting}.

\subsection{Calculation for open boundary conditions}
Consider an open square sample of side $L$, with two weak edges which meet at the corner situated at the origin. Thus, two perpendicular sets of 1D edge modes and one 0D corner mode are present, in addition to the bulk modes. The extension to situations with more than one corner mode is straightforward. To simplify the analyses, we will assume that the sample is large so that the number of bulk modes is much larger than the number of edge modes which is smaller by a factor $1/\sqrt{N}$. We will consider the superconducting OP at three specific locations as follows: 

--  corner site OP (site index $i=0$):
$\Delta_{corner}=V\langle c_{0\downarrow}c_{0\uparrow}\rangle$  

-- the OP at the center of the sample (site index $i=B$): 
$\Delta_{bulk}=V\langle c_{B\downarrow}c_{B\uparrow}\rangle$

-- the OP for a site at the midpoint of a weak edge (site index $i=E$):
$\Delta_{edge}=V\langle c_{E\downarrow}c_{E\uparrow}\rangle$ 

Let $\epsilon_\nu$ be the eigenvalues of the noninteracting Hamiltonian Eq.~\ref{sshham} and $\{\eta_\nu\}$ the eigenmodes. We will suppose that they are ordered such that $\nu=0,..,C-1$, denotes the $C$ corner modes (which for a finite sample can take the values 0, 1, 2, and 4), followed by the edge modes, and finally the 2D bulk modes. By diagonalization one obtains the transformation $U$ which takes one from the real space basis set $\{c_i\}$ to a new basis $\{\eta_\nu\}$, i.e. 

\begin{eqnarray}
c_j = \sum_{\nu} U_{i\nu}\eta_\nu  \qquad
\eta_\nu = \sum_{i} U^{-1}_{\nu i} c_i
\end{eqnarray}
where $U^{-1}=U^T$, the matrix $U$ being real.
The absence of translational symmetry makes it difficult to solve the coupled gap equations for OBC. However, with simplifications, some limiting cases are solvable, as shown below.

\bigskip

\noindent {\it System at Half-filling} 

In terms of the $U$ transformation matrix, one can write the order parameter for the midpoint at $T=0$, $\Delta_{bulk}$, as follows:
\begin{eqnarray}
\label{eq:Deltabulk}
\Delta_{bulk} &=& V \sum_{\nu} U^2_{B  \nu}b_{\nu\nu} 
\end{eqnarray}
where contributions of expectation values $b_{\mu\nu}$ for $\mu\neq\nu$ are neglected. In the small $r$ limit, the corner mode contribution and edge mode contributions can be dropped -- the former decays very fast and therefore is zero in the center of the sample, and the latter is very small because the edge band is far from the Fermi level. One has then
\begin{eqnarray}
\label{eq:Deltabulk2}
\Delta_{bulk} \approx \frac{V}{4N} \sum_{\nu} b_{\nu\nu} 
\end{eqnarray}
where the sum is over the bulk modes $\nu \geq 2L$. In the equation above, we have simplified by replacing the coefficients $U_{M \mu}^2$ by their average value $\overline{U}^2=1/4N$. To compute the $b_{\mu\mu}$, we assume that the interaction term can be decomposed into 2 by 2 blocks $H_\nu$ in the space $\{c_{\nu\downarrow},c^\dag_{\nu\uparrow}\}$, as in the periodic case.
In the case of bulk modes, the energies $\epsilon_\nu$ are essentially the same as the energies $\epsilon_{2D}$ in Eq.~\ref{eq:spectrum}. As a result, one obtains the same gap equation as in Eq.~\ref{eq:gapeqPBC}. In conclusion, $\Delta_{bulk} \approx \Delta$ and the bulk OP is essentially the same as the order parameter found for PBC.

One can proceed in a similar way for the corner site OP  $\Delta_{corner}$.  One finds 
\begin{eqnarray}
\label{eq:Deltacorner}
\Delta_{corner} &=& VU^2_{00} b_{00} + \sum_{\nu\in \mathrm{bulk}} U^2_{0\nu}b_{\nu\nu} + ... \nonumber \\
 &\approx& a_1V +  a_2  \Delta
\end{eqnarray}
where $\Delta$ represents the bulk OP given by Eq.~\ref{eq:gapeqPBC}. In the second line, the coefficients $U_{0\nu}^2$ have been replaced by their average value, written as $a_2/N$ with $a_2<1$ a constant of order 1.  In addition, we used the result of the BdG hamiltonian for the $\eta_0$ mode, which gives $b_{00}=1$. The constant  $a_1=U_{00}^2$. Both the coefficients $a_1$ and $a_2$ depend on the hopping ratio.

\noindent
When $r \rightarrow 0$, $a_1 \rightarrow \frac{1}{4}$ and $a_2\rightarrow 0$. Then $\Delta_{corner}=V/4$. Similarly, in this limit, the critical temperature for the transition at the corner can be shown to scale as $T_{c,corner}\sim V$. 

\noindent
When  $r\sim 1$, all coefficients of the $U$ matrix are of the same order of magnitude, $O(1/\sqrt{N})$. In this case, bulk modes contribute to leading order to all $\Delta_i$, while the corner and edge modes can be neglected. This results in bulk OP and corner OP of the same order of magnitude, and both are similar to $\Delta$ computed for the periodic case.  This explains the results shown in Fig.~\ref{deltaVfitting} for the corner and bulk order parameters as a function of $V$ for different hopping ratios $r$.  

\bigskip
\noindent {\it Chemical potential in an edge band.}  

When the Fermi level lies within an edge band, the superconducting gap opens within this band. The edge mode contributions are the most important and the OP is the largest on the edges. Writing out the expansion of $\Delta_{edge}$, one has 
\begin{eqnarray}
\Delta_{edge}= V\sum_\nu U_{E\nu}^2 \langle c_{\nu\downarrow}c_{\nu\uparrow}\rangle + ...
\end{eqnarray}
where $i$ is the index of the midpoint of a weak edge, and the sum runs over the indices $\mu=2,2L-1$. As before, we approximate the coefficients $U_{E\nu}^2$ by their average values. The equation can then be simplified to give the self-consistent equation for this OP at $T=0$ as follows
\begin{eqnarray}
\label{eq:edgeOP}
\Delta_{edge}\approx \frac{V}{4\sqrt{N}}\sum_\nu \frac{\Delta_{edge}}{\sqrt{(\epsilon_\nu-u_{HF\nu}-\mu)^2 + \Delta^2_{edge}}} \nonumber \\
\end{eqnarray}
In this expression, for small $r$ the single particle energies $\epsilon_\nu$ are essentially the 1 dimensional energies $\epsilon_{1D}$ written in Eq.~\ref{eq:spectrum}. Qualitatively, the above equation predicts that when the chemical potential lies within this band, the solution for the OP is expected to have the usual BCS form \footnote{Note however that when the chemical potential is in the region of the square root van Hove singularities, the dependence of the OP on $V$ should in principle be modified, with $\Delta_{edge} \sim \sqrt{V}$. However, the degeneracy of the edge bands (a factor 1/L smaller than that of the bulk) is too small for this effect to be of practical significance.}. The numerical results described in section III are in good accord with the analysis given here.

\section{Conclusions}
We have presented a detailed study of the inhomogeneous superconducting states found in the 2D extension of the Su-Schrieffer-Heeger model. We have focused on topologically non-trivial finite systems, where edge modes give rise to corner and edge superconducting phases.  The present model is of particular interest since it provides an analytically tractable example which can be solved in real space. We have discussed the band structure of the non-interacting model, and solutions in some simple limits for the interacting model treated in Bogoliubov-de Gennes mean field theory. Numerical solutions have been obtained for the full range of hopping ratio $t_1/t_2$. Depending on the hopping ratio and the band filling, we showed that the critical temperatures for these transitions scale in different ways with the Hubbard interaction $V$. The dependence can be linear or, vary as exponential-square-root or follow the standard BCS form.  

We have obtained the phase diagram of the superconducting phase and shown that an interesting mixed phase can occur above the bulk transition temperature $T_c$, in which the bulk can have a non-zero superconducting order induced by a proximity effect from the corners. 

Experimentally, 2D SSH lattices could be realized by the bottom-up assembly of atoms, a method which has been to fabricate ``designer" structures, as described in \cite{khajetoorians2019creating}. The superconducting states, in particular, for the edge and corner superconducting phases, could be probed by scanning tunnelling spectroscopy (STS). 
An interesting direction for future work consists of extending the length of the unit cell of the 1D chains used in defining the 2D model. One can get edge and corner modes in 2D systems by considering chains of period 3 -- following a $\{t_1t_2t_1\}$ sequence in $x$ and $y$ directions. This is a member of a set of finite sequences which in the infinite limit give rise to the Fibonacci quasicrystal, known to host topological edge modes \cite{kraus2012topological,rai2021bulk}. A recent paper has introduced a different route towards a 2D topological model starting from SSH chains \cite{rosenberg2022}. The parameters of the model they considered are different, and boundary phenomena were not discussed, however, and it would be interesting to compare their model predictions with ours for finite samples. It is not difficult to generalize our model to 3D, by taking a direct product of three orthogonal SSH chains, in which case, vertex edge, surface and bulk modes should appear.  Last but not least, Floquet topological systems such as those discussed in \cite{nag2021hierarchy,ghosh2021hierarchy} constitute another class of systems likely to host interesting edge and corner superconducting states. 

Future work will involve going beyond mean field theory to investigate the stability of the low dimensional phases described in this work. It will be interesting to see how the results are modified when fluctuations are included. It would be informative and interesting to observe the edge and corner superconductivity experimentally. The experimental realization of a 2D SSH lattice should be straightforward for the non-interacting limit. The attractive Hubbard Hamiltonian is harder to realize but could be envisaged in cold atoms systems, for example. In this work, we have considered only $s$-wave superconductivity. Adding spin-orbit interactions should lead to new interesting edge phenomena and competition between order parameters of different pairing symmetries close to the edges, as discussed in \cite{lauke2018friedel}.

\section{Acknowledgement}
The authors acknowledge the Center for Advanced Research Computing (CARC) at the University of Southern California for providing computing resources that have contributed to the research results reported within this publication. URL: https://carc.usc.edu.
%
%
%
\bibliographystyle{unsrt}
\bibliography{biblio.bib}

\begin{thebibliography}{10}

\bibitem{su1979solitons}
W\_P Su, JR~Schrieffer, and Ao~J Heeger.
\newblock Solitons in polyacetylene.
\newblock {\em Physical review letters}, 42(25):1698, 1979.

\bibitem{su1983erratum}
WP~Su, JR~Schrieffer, and AJ~Heeger.
\newblock Erratum: Soliton excitations in polyacetylene.
\newblock {\em Physical Review B}, 28(2):1138, 1983.

\bibitem{trifunovic2021higher}
Luka Trifunovic and Piet~W Brouwer.
\newblock Higher-order topological band structures.
\newblock {\em physica status solidi (b)}, 258(1):2000090, 2021.

\bibitem{benalcazar2017electric}
Wladimir~A Benalcazar, B~Andrei Bernevig, and Taylor~L Hughes.
\newblock Electric multipole moments, topological multipole moment pumping, and
  chiral hinge states in crystalline insulators.
\newblock {\em Physical Review B}, 96(24):245115, 2017.

\bibitem{benalcazar2017quantized}
Wladimir~A Benalcazar, B~Andrei Bernevig, and Taylor~L Hughes.
\newblock Quantized electric multipole insulators.
\newblock {\em Science}, 357(6346):61--66, 2017.

\bibitem{xu2020general}
Xun-Wei Xu, Yu-Zeng Li, Zheng-Fang Liu, and Ai-Xi Chen.
\newblock General bounded corner states in the two-dimensional
  su-schrieffer-heeger model with intracellular next-nearest-neighbor hopping.
\newblock {\em Physical Review A}, 101(6):063839, 2020.

\bibitem{geier2020symmetry}
Max Geier, Piet~W Brouwer, and Luka Trifunovic.
\newblock Symmetry-based indicators for topological bogoliubov--de gennes
  hamiltonians.
\newblock {\em Physical Review B}, 101(24):245128, 2020.

\bibitem{luo2022higher}
Xun-Jiang Luo, Xiao-Hong Pan, Chao-Xing Liu, and Xin Liu.
\newblock Higher-order topological phases emerging from the
  su-schrieffer-heeger stacking.
\newblock {\em arXiv preprint arXiv:2202.13848}, 2022.

\bibitem{scammell2022intrinsic}
Harley~D Scammell, Julian Ingham, Max Geier, and Tommy Li.
\newblock Intrinsic first-and higher-order topological superconductivity in a
  doped topological insulator.
\newblock {\em Physical Review B}, 105(19):195149, 2022.

\bibitem{li2020artificial}
Tommy Li, Julian Ingham, and Harley~D Scammell.
\newblock Artificial graphene: Unconventional superconductivity in a honeycomb
  superlattice.
\newblock {\em Physical Review Research}, 2(4):043155, 2020.

\bibitem{padavic2018topological}
Karmela Padavi{\'c}, Suraj~S Hegde, Wade DeGottardi, and Smitha Vishveshwara.
\newblock Topological phases, edge modes, and the hofstadter butterfly in
  coupled su-schrieffer-heeger systems.
\newblock {\em Physical Review B}, 98(2):024205, 2018.

\bibitem{ghosal1998role}
Amit Ghosal, Mohit Randeria, and Nandini Trivedi.
\newblock Role of spatial amplitude fluctuations in highly disordered s-wave
  superconductors.
\newblock {\em Physical review letters}, 81(18):3940, 1998.

\bibitem{aryanpour2006effect}
K~Aryanpour, Elbio~R Dagotto, Matthias Mayr, T~Paiva, WE~Pickett, and Richard~T
  Scalettar.
\newblock Effect of inhomogeneity on s-wave superconductivity in the attractive
  hubbard model.
\newblock {\em Physical Review B}, 73(10):104518, 2006.

\bibitem{hirsch1985two}
Jorge~E Hirsch.
\newblock Two-dimensional hubbard model: Numerical simulation study.
\newblock {\em Physical Review B}, 31(7):4403, 1985.

\bibitem{noda2015bcs}
Kazuto Noda, Kensuke Inaba, and Makoto Yamashita.
\newblock Bcs superconducting transitions in lattice fermions.
\newblock {\em arXiv preprint arXiv:1512.07858}, 2015.

\bibitem{noda2015magnetism}
Kazuto Noda, Kensuke Inaba, and Makoto Yamashita.
\newblock Magnetism in the three-dimensional layered lieb lattice: Enhanced
  transition temperature via flat-band and van hove singularities.
\newblock {\em Physical Review A}, 91(6):063610, 2015.

\bibitem{samoilenka2020boundary}
Albert Samoilenka and Egor Babaev.
\newblock Boundary states with elevated critical temperatures in
  bardeen-cooper-schrieffer superconductors.
\newblock {\em Physical Review B}, 101(13):134512, 2020.

\bibitem{samoilenka2020pair}
Albert Samoilenka, Mats Barkman, Andrea Benfenati, and Egor Babaev.
\newblock Pair-density-wave superconductivity of faces, edges, and vertices in
  systems with imbalanced fermions.
\newblock {\em Physical Review B}, 101(5):054506, 2020.

\bibitem{benfenati2021boundary}
Andrea Benfenati, Albert Samoilenka, and Egor Babaev.
\newblock Boundary effects in two-band superconductors.
\newblock {\em Physical Review B}, 103(14):144512, 2021.

\bibitem{Note1}
Note however that when the chemical potential is in the region of the square
  root van Hove singularities, the dependence of the OP on $V$ should in
  principle be modified, with $\Delta _{edge} \sim \protect \sqrt {V}$.
  However, the degeneracy of the edge bands (a factor 1/L smaller than that of
  the bulk) is too small for this effect to be of practical significance.

\bibitem{khajetoorians2019creating}
Alexander~A Khajetoorians, Daniel Wegner, Alexander~F Otte, and Ingmar Swart.
\newblock Creating designer quantum states of matter atom-by-atom.
\newblock {\em Nature Reviews Physics}, 1(12):703--715, 2019.

\bibitem{kraus2012topological}
Yaacov~E Kraus, Yoav Lahini, Zohar Ringel, Mor Verbin, and Oded Zilberberg.
\newblock Topological states and adiabatic pumping in quasicrystals.
\newblock {\em Physical review letters}, 109(10):106402, 2012.

\bibitem{rai2021bulk}
Gautam Rai, Henning Schl{\"o}mer, Chris Matsumura, Stephan Haas, and Anuradha
  Jagannathan.
\newblock Bulk topological signatures of a quasicrystal.
\newblock {\em Physical Review B}, 104(18):184202, 2021.

\bibitem{rosenberg2022}
Peter Rosenberg and Efstratios Manousakis.
\newblock Topological superconductivity in a two-dimensional weyl ssh model.
\newblock {\em Phys. Rev. B}, 106:054511, Aug 2022.

\bibitem{nag2021hierarchy}
Tanay Nag, Vladimir Juri{\v{c}}i{\'c}, and Bitan Roy.
\newblock Hierarchy of higher-order floquet topological phases in three
  dimensions.
\newblock {\em Physical Review B}, 103(11):115308, 2021.

\bibitem{ghosh2021hierarchy}
Arnob~Kumar Ghosh, Tanay Nag, and Arijit Saha.
\newblock Hierarchy of higher-order topological superconductors in three
  dimensions.
\newblock {\em Physical Review B}, 104(13):134508, 2021.

\bibitem{lauke2018friedel}
Lars Lauke, Mathias~S Scheurer, Andreas Poenicke, and J{\"o}rg Schmalian.
\newblock Friedel oscillations and majorana zero modes in inhomogeneous
  superconductors.
\newblock {\em Physical Review B}, 98(13):134502, 2018.

\end{thebibliography}
\end{document}